\documentclass[acmsmall]{acmart}

\AtBeginDocument{%
  \providecommand\BibTeX{{%
    \normalfont B\kern-0.5em{\scshape i\kern-0.25em b}\kern-0.8em\TeX}}}

\setcopyright{acmlicensed}
\acmJournal{PACMHCI}
\acmYear{2022} \acmVolume{6} \acmNumber{CSCW1} \acmArticle{94} \acmMonth{4} \acmPrice{15.00}\acmDOI{10.1145/3512941}

\newcommand\change[1]{{\color{black} #1}}
\newcommand\changege[1]{{\color{black} #1}}

\begin{document}

\title[How Employees Collaborate Through Multi-User Chat Channels on Slack]{Group Chat Ecology in Enterprise Instant Messaging: How Employees Collaborate Through Multi-User Chat Channels on Slack}

\author{Dakuo Wang}
\email{dakuo.wang@ibm.com}
\affiliation{%
 \institution{IBM Research}
  \city{Cambridge}
  \state{MA}
  \postcode{02142}
  \country{USA}
}

\author{Haoyu Wang}
\authornotemark[1]
\affiliation{%
  \institution{Amazon Alexa}
  \country{USA}
 }

\author{Mo Yu}
\authornotemark[1]
\affiliation{%
  \institution{WeChat AI, Tencent}
  \country{USA}
}

\author{Zahra Ashktorab}
\affiliation{%
  \institution{IBM T.J. Watson Research Center}
  \country{USA}
}

\author{Ming Tan}
\affiliation{%
  \institution{Amazon}
  \country{USA}
}
\authornote{This work was done while Haoyu, Mo and Ming worked at IBM.}

\renewcommand{\shortauthors}{Dakuo Wang et al.}

\begin{abstract}
  Despite the long history of studying instant messaging usage, we know very little about how \change{today's} people participate in group chat channels and interact with others \change{inside a real-world organization}. In this short paper, we aim to update the existing knowledge on how group chat is used in the context of today's organizations. The knowledge is particularly important for the new norm of remote works under the COVID-19 pandemic. \change{We have the privilege of collecting two valuable datasets: a total of 4,300 group chat channels in Slack from an R\&D department in a multinational IT company; and a total of 117 groups' performance data.} Through qualitative coding of 100 randomly sampled group channels from the 4,300 channels dataset, we identified and reported \textbf{9 categories} such as \textit{Project} channels, \textit{IT-Support} channels, and \textit{Event} channels. We further defined a feature metric \change{with 21 meta features (and their derived features) without looking at the message content} to depict the group communication style for these group chat channels, with which we successfully trained a  machine learning model that can automatically classify a given group channel into one of the 9 categories. 
  In addition to the descriptive data analysis, we illustrated how these communication metrics can be used to analyze team performance. 
  \change{We cross-referenced 117 project teams and their team-based Slack channels and identified 57 teams that appeared in both datasets, then we built a regression model to reveal the relationship between these group communication styles and the project team performance.
  This work contributes an updated empirical understanding of human-human communication practices within the enterprise setting, and suggests design opportunities for the future of human-AI communication experience.} 
\end{abstract}


%

\maketitle

\section{Introduction and Background}
Instant messaging systems (IM) emerged in 1980s then slowly being adopted by individuals as well as organizations in the past few decades. With years of research and development effort in CSCW (e.g.,~\cite{bradner1999adoption, erickson1999socially, nardi2000interaction,dourish1992awareness, olson2000distance, mcdaniel1996identifying,hinds2003out,whiting2019did}), modern IMs (e.g., Slack and Skype) are much more advanced. More and more companies and organizations are taking these IMs as granted that their workers exchange information and files so efficiently. \change{Slack is one such tool that has been adopted by offices and work spaces~\cite{cao2021my,lin2016developers,stray2019slack,stray2020understanding,chatterjee2019exploratory,chatterjee2020software}.} Comparing to the previous generation of IMs used in workplaces (e.g, Lotus Sametime), it emphasizes more on the group chat feature and provides a much better user experiences for multi-parties chats. It is estimated that Slack reduces emails by 32\% and reduce meetings by 23\%, it also helps new employees to reach full productivity 25\% sooner~\cite{slackreport}.

In parallel to the recent advancement of IM technologies, text-based AI applications (e.g., Chatbots or topic summarization) reside in IMs has attracted many interests from researchers and business. These applications are often built based on a language model that can simulate how human chat with another human~\cite{shamekhi2018face,wang2020human,wang2021cass}. Many HCI researchers and natural language processing (NLP) researchers have invested tons of efforts in building better algorithms and applications, but often failed to do so \cite{boutin_boutin_2017,fan2021utilization}. Partially because many of such efforts have been based on an \change{outdated} understanding of how people chat with each other and the communicate styles ( e.g., ~\cite{wang2019extracting} and many other NLP works assumed that a conversation thread has more than 10 turns and less than 100 turns). Thus, there is an urgent need to update the existing understanding of how people are chatting in groups and in a finer-grained level, such as how many conversation threads with different topics are incurred in a group chat channel and how deep each thread is.

Another dimension of fine-grain understanding is the group chat ecosystem. Often time AI researchers focus only on a single context of group chats~\cite{wang2019extracting}. For example, the famous human-human group chat datasets being used by NLP researcher are Ubuntu-IRC ~\cite{elsner2010disentangling} and the Twitch dataset ~\cite{hamilton2014streaming}, which focus on IT development and Gaming thus has limited generalizability. Another line of research relies on synthetic datasets often generated from the Reddit~\cite{wang2019extracting} or StackOverflow, but these synthetic datasets from online forums do not guarantee the similarity in communication styles in the real IMs datasets. 

\changege{This paper aims to \textbf{1) update the understanding of organizational group chat practices, and 2) examine whether the group chat styles can quantitatively reflect these groups' team performance}
. } 
\changege{As a first step,} we collected a total number of 4,300 group chat channels being created and used by ~8,000 employees in a R\&D division in a big IT company. We collected both the meta data and the raw messages of these channels spanning from Mar 2016 (Slack was first introduced) to Mar 2019 (this Paper was written). Then we randomly selected 100 group channels and manually coded them into 9 different categories. \change{One particular challenge is that the content of such organizational data corpus is often proprietary, thus we further designed a machine learning classification model that leverages only the meta data of such chat channels (e.g., \# of messages and \# of threads), but not the content, to identify a channel's category. }

\changege{As for the second research goal,} \change{many prior literature have suspected that team's communication and collaboration patterns may predict team performance~\cite{cao2021my,fussell1998coordination,howell2006effects,olson2017people,wang2021organizational}. In our study, we have the privilege to access another data corpus with 117 teams' performance data. Through cross-referencing these two datasets, we identified 54 teams that existed in both data corpus. \changege{Thus, our paper can also examine whether a group's communication style can quantitatively reflect these groups' team performance.}}

\section{Related Work}
\subsection{Instant Messaging Systems and Research in CSCW}
There are decades of CSCW literature on IM system designs and user studies of it that we do not have space to cover all of them. We only want to point out one particular genre of research on IMs that is around multiple-parties group chat. Back in 1994, McDaniel et al. compared how people chat with each other in Face-to-Face setting (FTF) versus in text-based Computer-Medicated Communication (CMC) systems, which is the early name of IMs and Videoconferencing systems~\cite{mcdaniel1996identifying}. They found that a group of people chat on multiple concurrent threads in a conversation (2 to 3 for FTF and 4 to 6 for CMC) and the threads have a different timespan (2.8 mins for FTF and 23.3 mins for CMC). These results are intriguing but the data corpus and the analysis method they used were quit primitive: they analyzed 6 chatting groups and labeled the timestamp for each message, put them on the same timeline, and manually counted the threads and the numbers.

Another notable research work about early days IMs usage is from Bonnie Nardi and Steve Whittaker ~\cite{nardi2000interaction} in 2000. They particularly focused on the early adopters of an IM system in a workplace setting and reported how they used or would like to use IMs. For example, they reported that users use IMs not only for work-related question and answering type of activities ("Interaction"), they also use it for informal purposes such as checking whether someone is available for a chat or not. And they also reported that those early-days IM systems were designed based on a dyadic "call" model, where users use it more like a phone to find another individual to chat, but preferred not to use the "chat room" type of features.

Together with other research effort, the design implications stemmed from these research findings have guided the following two decades of IM system design, e.g., having a status feature and lacking a groupchat feature in workplace IMs. Now that new tools exist and people are using them, it is time for us to revisit this research topic after ~20 years.

\subsection{Communication Style Recognition in HCI and in NLP}
In addition to the HCI research effort of understanding human chat behavior, NLP researchers have also investigated how to advance techniques to automatically analyze text-based human conversations.
However, most of these NLP researches focus on the one-on-one conversation scenarios (i.e. conversation between a chatbot and a human user), on which information retrieval traditional methods, syntactic/semantic parsing techniques, and neural sequence-to-sequence generation models are integrated into the chatbot \cite{higashinaka2014towards,yan2016docchat,zhou2018design}.
For the domain of analyzing group chat conversations, there are a few works on disentangling interleaved conversational threads to form threads discussing single topics~\cite{wang2019extracting,elsner2010disentangling, mayfield2012hierarchical,jiang2018learning} and extracting knowledge from conversational dialogues~\cite{hixon2015learning}.
These works have limitations in the sense that they neglect the richness in conversation patterns in different conversational categories.
For example, the work~\cite{wang2019extracting} was conducted on particular interest-based Reddit forums; and the work~\cite{hixon2015learning} focused only on educational related topics, thus they have limited generalizability to other domains or contexts.

\change{One notable research work that is particularly relevant to our project is from Lin et al.~\cite{lin2016developers} They designed an exploratory study to investigate how software engineers use Slack. Through a survey study, the survey respondents replied why and how they use Slack. For example, the respondents mentioned three types of benefits of using Slack: personal benefits, team-wide benefits, and community benefits. Lin's work sheds light on the variate use patterns and perceived benefits of group chat, but because their data were only from qualitative survey results, it did not provide actionable insights on how to materialize their findings into the design of algorithms or applications in the group chat scenario.

Our work focuses on the real-world workplace group chat ecology. We argue a better categorization of different types of conversation groups through modeling with these chat groups' meta-information (without looking at the potentially sensitive message content) is necessary for the downstream NLP algorithms or HCI system building tasks.}

\subsection{Communication Style and Team Performance}

\changege{One downstream application of identifying communication styles for groups is that we may use it to predict the team performance.} There has been extensive prior literature on this topic. For example, Zhang et al. applied topic models analysis on chat messages to investigate the evolution of team dynamics over a long term project \cite{zhang2018team}. They describe common behaviors and team cohesion dynamics. 
It is well known in the CSCW community that people who work on teams may not put as much effort than if they were working individually (i.e., "Slackers") \cite{karau1993social}. Furthermore, coordinating individuals' contributions through communication is challenging and critical \cite{diehl1987productivity}, and many CSCW systems have been proposed to address some of these challenges \cite{kraut2003applying}. 

\change{One notable recent work from Cao et al.~\cite{cao2021my}. They designed an online experiment and recruited crowdworkers to form chat groups and to perform a group-based task. Then, with various machine-generated features and human-labeled features, they built machine learning models to predict team's viability, which is a metric reflecting how successful the team collaboration is. Their machine-generated features include mostly the computational linguistic features (e.g., pronoun) derived from the message content and two meta features (i.e., \# of messages per person, and \# of words per person).}

In this current work, we follow the trend of adopting machine learning methods to extract group behaviors and then to predict the group performance. \change{In comparison to these prior works, our work investigates a real-world dataset as opposed to an experimental study (e.g., ~\cite{cao2021my}), and our model does not require the actual message content to preserve user privacy. }

\section{Methods} 
In this section, we describe our datasets, the open coding analysis method we used to identify the 9 categories of group chat channels, the machine learning methods that we used to pre-process the datasets and extract the feature metrics, and the regression method that we used to analyze the relationship between the features metrics and the team performance in a subset of 54 project team's dataset.

\subsection{Datasets}
In total, there are two datasets being used in this study, a Slack message dataset with 4,300 public channels in an R\&D department in a company, and a dataset with 117 project teams with the team composition and performance information. 

For the 4,300 channels dataset, we randomly select 100 channels as a subset for further manual coding analysis. For the 117 project team performance dataset, we cross-reference with the 4,300 project channels and identified that 54 project teams have a designated and \textbf{publicly-accessible} group chat channel, thus we then prepared a sub-dataset for the 54 project teams with both the project performance and its Slack group chat data.

Here, we would like to provide a bit more details about the Project Team performance dataset. In this R\&D department in the multinational IT company, a re-organization occurred in November 2017 and 117 project teams (not the same as organization teams) were formed. For a few months period after November 2017, these teams all were encouraged to submit papers to an academic conference as their primary goal, thus 146 submissions were generated by the conference submission deadline.

From an internal project management portal, we collected information about these project teams, such as the project description and team members' information. We use whether a project team generates a paper submission as the final outcome to reflect its performance. If there are one or more submissions to the conference, we denote 1 to the outcome variable, otherwise 0. 

We acknowledge that this way of describing team performance has many limitations and we will elaborate in the limitation section by the end of the paper. In addition to the project performance, each of the project team is also required to have a Slack group chat channel, but many of those channels are private to the team members due to confidentiality purpose. At the end, we are only able to collect 54 Slack channels for the 117 groups. 

\subsection{Manual Coding Slack Group Chat Channels to Identify Categories}
\change{Inspired by prior literature~\cite{lin2016developers} and based on our observation,} the group chat channels can vary tremendously in the number of members or some other characteristics. Thus, we decided to first conduct a qualitative analysis to identify the different types of Slack groups. We randomly selected 100 channels for manually labeling the categories, and we were prepared to code more channels if new categories keep emerging. The result of the 9 categories suggested that our code has reached the saturated thus we stopped. In particular, two authors of this paper independently conducted thematic content analysis for each of the 100 Slack channels by reading the content of the Slack group, and various meta-data such as channel description and the number of members. Independently the two authors coded each of the Slack groups and took notes why they believed so. Then, the two authors discussed their notes and coding schema (without revealing the code for each channel), and finalized a list of 9 categories (see Table \ref{tab:chall}). The two authors then re-coded the 100 slack channels with the agreed code list for a cohen kappa score of 0.83.

\begin{table}[htp]
\small
  \centering
  \caption{The 9 Categories of 100 Randomly Selected Slack Group Chat Channels Identified by Manual Coding, Each with a Short Description}
  \label{tab:chall}
  \begin{tabular}{p{3.5cm}|p{0.5cm}|p{7cm} }
    \toprule
    \textbf{Channel Category} & \textbf{N} & \textbf{Description}   \\
    \midrule
    Project &32 & This category of channels consists with discussions around projects   \\
	Social Group &10& This category involves discussions on non-work related social activities   \\
	IT Support &8& This category of channels often being used as a help desk for internal systems or tools  \\
	Employee Support& 3& The channels being used as a help desk for answer HR or other logistics related questions   \\
	Tech Enthusiasts& 10& Often consists with a group of members discussing a new technology that are not necessarily related to their work project \\
	Event &9& Schedule, plan and discussion around a temporary event at work \\
	Bot& 8& A type of channels where most of the messages are from a Slack bot, often being used to monitor a particular system's maintenance log\\
	Test &18& Users mistakenly create a group channel or tested the creation of channel function, often no message is posted   \\
	Announcement& 2 & A channel for department or even organizational level announcement\\
  \bottomrule
  \end{tabular}
\end{table}

\subsection{Feature Metrics to Represent Communication Styles of a Slack Group Chat Channel}
\change{To depict the group chat ecology and to support our research purpose, we collected 21 features based on its meta information to represent a group chat channel. These 21 features (summarized in Table~\ref{tab:feat}) are the whole suite of features that Slack API can provide to a developer or researcher without accessing the actual message content. }

Many of these features are self-explanatory thus here we focus only on the confusing ones to elaborate.

The \textit{\#members} is the number of people who have joined in the channel and \textit{\#active\_users} represents the number of people with at least one message in the channel. Active members could be larger than current members because people may leave the channel after they post messages. 

\textit{\#active\_timespan} captures the number of days from the day that the first message was posted in the channel to the day that the last message was posted. \textit{word\_per\_message} is an average number words in each message. Sometimes there might be a single user dominated the whole group channel that we also measure the max number of messages by one user.

A very common interactive function in Slack is to notify a specific user by \textit{@username} or the entire group by \textit{@channel} or \textit{@here}. We thus include the corresponded number of messages for those three different types of messages as \textit{\#at\_messages}, \textit{\#channel\_messages}, and \textit{\#here\_messages}. A more complicated interaction is that many people tend to explicitly form a ``\textbf{thread}'' in a Slack channel for a small discussion on one topic, we generate feature \textit{\#threads} to represent the number of threads within the channel. We also have \textit{max\_turn\_thread} and \textit{avg\_turn\_thread} to capture the maximum and average number of turns in a thread. Another common used function in Slack is to ``\textbf{react}'' to a specific message by a simple emoji. We have feature \textit{max\_count\_reaction} and \textit{avg\_count\_reaction} for maximum and average number of reactions for a message.

\textit{\#emoji\_messages} will capture the number of messages with emojis within it. The \textit{\#pinned\_messages} represents the number of messages ``\textbf{pinned}'' by the users in the channel which they feel is important. Since people may share code snippets in a channel, we introduce \textit{\#code\_messages} for the number of messages with code snippets. 

We also have \textit{\#url\_messages} and \textit{\#git\_messages} to represent the number of messages which contains an URL and the number of messages which contains an URL specific to a Github page. People are also able to easily share files within a channel, we also capture such behavior by having \textit{\#file\_messages} to capture the number of messages contains file sharing. As for Slack, the owner of the channel could introduce a ``\textbf{Slack bot}'' which interacts with people in different ways such as general question answering bot, alert notification bot, etc. We also count the number of messages generated by the Slack bots as \textit{\#bot\_messages}.

\change{Inspired by liteature~\cite{cao2021my}, some of these features may highly correlated to how many members are there in a chat channel, how long the channel has been established, and how many total messages there are. Thus, for each of the count-based features (from \textit{feature 6 }to \textit{feature 21}), we also compute three normalized derived features by dividing \textit{\#messages} (producing 12 new features), \textit{\#active\_users} (producing 13 new features) and \textit{active\_timespan} (producing 13 new features). In total, we end up having 59 meta features as representation of each slack group in the machine learning model.}

\begin{table}[htp]
\small
  \centering
  \caption{List of 21 Features of Slack Group Channels Meta Information from Three Representative Categories. The full table is included in Appendix.}
  \label{tab:feat}
  \begin{tabular}{p{0.5cm}|p{2.5cm}|p{5.7cm} |p{1cm}|p{1cm}|p{1.3cm}}
    \toprule
   \textbf{ID} & \textbf{Feature} & \textbf{Definition} & \textbf{Project N=32} & \textbf{Social Group N=10} & \textbf{IT Support N=8}\\
    \midrule
 1& \#current\_members & \#current members in the channel & 11.1 & 84.0 & 177.0\\
 2& \#active\_members & \#members with at least one message & 14.4 & 92.7 & 139.9\\
3& active\_timespan & \#days between first message and last message & 281.3 & 478.6 & 506.5\\
4& \#message\_top\_user & the max \#messages posted by one user & 90.2 & 554.9 & 1819.9\\
5& word\_per\_message & avg length of message in \#words & 16.4 & 12.8 & 23.0\\
6& \#messages & \#messages in total in this Slack channel & 355.9 & 3059.0 & 4537.4\\
7& \#at\_messages & \#messages with @user mentions & 79.1 & 337.0 & 953.3\\
8& \#channel\_messages & \#messages with @channel broadcast mention & 0.0 & 0.1 & 0.1\\
9& \#here\_messages & \#messages with @here broadcast mention & 0.1 & 8.1 & 1.6\\ 
10& \#threads\_messages & \#threads in the channel & 15.3 & 152.0 & 185.8\\
11& max\_turn\_thread & max \#messages in a thread & 6.7 & 28.6 & 34.9\\
12& avg\_turn\_thread & avg \#messages in a thread & 2.0 & 2.18 & 4.1\\
13& \#reacted\_messages & \#messages with reaction emoji & 19.5  & 549.0 & 155.0\\
14& avg\_count\_reaction\ & avg \#reactions in a reacted message & 0.9 & 2.2 & 1.5 \\
15& \#emoji\_messages & \#messages with emoji & 11.9 & 284.7 & 102.4\\
16& \#pinned\_messages & \#pinned messages & 0.44 & 1.9 & 5.5\\
17& \#code\_messages & \#messages with code segments & 6.5 & 1.4 & 81.8\\
18& \#url\_messages & \#messages with URL & 34.2 & 188.0 & 329.5\\ 
19& \#git\_messages & \#messages with Github URL & 8.7 & 0.7 & 59.6\\
20& \#file\_messages & \#messages with Files & 60.4 & 180.5 & 58.9\\
21& \#bot\_messages & \#messages sent by Slack chatbot & 52.1 & 64.6 & 1396.5\\

  \bottomrule
  \end{tabular}
\end{table}

\subsection{Machine Learning Model to Automatically Identify Categories}
\change{In order to provide an actionable research asset for fellow researchers and developers to classify group chat categories,} we conduct an automatic Slack group classification task using all 59 features. After extracting the feature vector for each labeled Slack channel, we build an ensemble tree-based classifier from \cite{geurts2006extremely} to predict the category of each channel. The rationales of choosing this model are two folds: first, given our features, it is straightforward for a decision-tree based algorithm to learn good rules that are non-linear combinations of features; second, the number of coded Slack groups are limited and an ensemble-based approach could help prevent over-fitting.

As for the evaluation process, we are focusing on the overall classification Accuracy as well as Precision and Recall for each label. Due to the limited coded channels, we follow the leaving-one-out cross-validation method from \cite{kearns1999algorithmic} to measure the overall classification accuracy, which is widely used for model evaluation on small data set.

\subsection{Recursive Logistic Regression Model Method to Examine the Relationship between a Project Team's Communication Style in Slack and Team Performance}

\changege{To fulfill our second research goal, we conducted a logistic regression on the subset of the data with 54 project teams. }\change{We use their Slack channel's features (all 59 with both raw features and derived features) as independent variables and the binary dependent variable representing whether they have a publication or not. Among the 54 project teams in our dataset, 35 had published papers and 19 had not published.
We used recursive feature elimination algorithm \cite{yan2015feature} to identify the best features in the model that were the most predictive.}

\section{Results}

\subsection{Communication Styles for the 9 Categories Based on Group Chat's Meta Data} 
Through the qualitative coding of 100 randomly sampled channels, we were able to identify 9 different categories. The code names and the descriptions are all as listed in Table~\ref{tab:chall}, thus we would not repeat here.

Based on the manual coding of 100 data points, the machine learning algorithm \cite{geurts2006extremely} can also pick up the difference between categories. Through our experiment, the machine learning model for identifying different categories could achieve 66\% overall Accuracy on the coded 100 channels. As for \textit{Project} category, we could get Precision 79.4\% and Recall 87.1\%, which we believe is reasonable good for our downstream task (Section \ref{exp2}) on this specific category. We could also get reasonable performance for other labels except for \textit{Employee Support} and \textit{Announcement}, simply because we have only 2-3 data points for each of those categories.

By examining the features for each category, we notice that there are several different communication styles. 
Within table \ref{tab:feat}, we provide the average value for each feature of three different categories\footnote{\change{We listed only the three categories here as \textit{Project} is the predominant type of channels, Tech Enthusiasts and Social Group are relatively similar, IT Support and Employee Support are also similar, the rest of channels do not have much group interaction (e.g., Announcement, Bot, and Test). For a full list of descriptive statistics, please refer to Appendix Table \ref{tab:feat_full}}} including \textit{Project}, \textit{Social}, and \textit{IT Support} . It is quite intuitive to see the differences in some of characteristics of each category by looking at several representative features: 

\begin{itemize}
\item \textbf{Project Category:}
From the results in table \ref{tab:feat}, we can find that a \textit{Project} slack channel usually has fewer messages~(355.9) and fewer number of members~(11.1) compared to the other two groups which have more than 3,000 messages on average and about 100 members. We believe this is natural for a \textit{Project} Slack channel, since it consists of a few people working on the project that form a centralized communication. Even the total number of messages is limited as they may communicate locally as well, the percentage of \textit{\#file\_messages} are substantially higher than the other categories as we believe members in such channels tend to share files for productive collaboration on a project.

If we examine \textit{active\_timespan} of this category, we could also notice that the active time span is shorter than the other two categories because of a project is supposed to finish within a period. We also notice similar characteristic for \textit{Event} group where the active time span is even shorter (144 days), as people tend to quickly form a slack group discussion for a specific event and then become inactive as the event finishes.

\item \textbf{Social Group Category:}
A \textit{Social} group usually has a large number of messages, where the amount of messages is similar comparing to the \textit{IT Support} group we will discuss below.
But the \textit{Social} groups could be differentiated from the other types of groups, including the \textit{IT Support} groups, based on features like \textit{\#emoji\_messages} and \textit{\#reacted\_messages}.

For this category, we could observe more frequent usages of emojis~(284.7) because of the more casual style communication in such channels. Similarly, we notice that for \textit{Bot} channels, the emojis are also widely used as people tend to build the Slack bots with emojis in the conversation to make the bots more friendly. For the \textit{Social} groups, another feature with higher values than the others is the \textit{\#reacted\_messages}, which is the number of messages with emoji reactions like thumbs-up and thumbs-down. This is similar to the number of emoji messages as people tend to form a casual communication style. Members in this channel also tend to use \textit{@here} more frequently than in other channels. We hypothesis the reason behind is that people wish to eagerly share content with all other members.

\item \textbf{IT Support Category:} As for \textit{IT Support} category, it usually involves messages trying to solve a specific technical issue so that the number of messages containing code snippet~(81.8) is significantly higher than the other two groups which have less than 10 such messages.

There are some other features which could help differentiated this category. We notice that the averaged number of turns in threaded messages~(4.1) for this category is significantly higher than all the other categories. We believe the reason is that people need multiple turns of conversation in order to solve an technical issue. If we look at the feature \textit{\#message\_top\_user}, we could notice that the value for \textit{IT Support}~(1819.9) is substantially higher than others and we think it suggests that the user behind is the one who actively provides solutions to most of the IT problems. As for \textit{\#bot\_messages}, this type of channel has more messages sent by Slack bot. This is in accordance with our findings that many \textit{IT Support} channels are using Slack bots to handle some basic questions and frequently asked questions (FAQ) with regard to service status or product update. We also notice a significant higher usage~(953.3) of \textit{@username} and we believe it is for people to tag specific users to solve a specific technical issue.
\end{itemize}

\subsection{\change{What Features in Group Communication May Reflect Team Performance}}
\label{exp2}
\change{To accomplish the second research goal of assessing the feasibility of modeling team performance with only the group communication meta data, we feed all the 59 features (raw and derived) in a recursive logistic regression model. The model selects the following final features for predicting whether the team has a submission or not}: active timespan in days, number of bot messages, number of @here messages, normalized messages for the top user, normalized threads, normalized emoji, and normalized code. These features yielded an $R^2$ of 0.61. The results of our logistic regression are in Table \ref{tab:reg}. Below we briefly discuss these features. 

\begin{table}[htp]
\small
  \centering
  \caption{ Logistic Regression Result Of 7 Features Lead to Team Performance }
  \label{tab:reg}
  \begin{tabular}{lllll}
    \textbf{Feature} & \textbf{B} & \textbf{S.E} & \textbf{z} & \textbf{Pr>|z|}  \\
    \midrule
  Active Time Span & 0.0023 & 0.002&1.446&0.148\\
  Number of Bot Messages &20.9500& 5.04e+04&0&1.0\\
  Normalized Messages Per Top User & 21.3561& 6.07e+04 &0&1.0\\
  Normalized Threads & 0.0099&0.011& 0.891&0.373\\
  Normalized Code &-1.2670 &  1.442 &-0.879&0.380 \\
  Normalized Emoji &-0.1345 & 0.228&-0.589&0.556\\
  Normalized @Here & -1.4987 &  5.136&0.292&0.770\\
    
  \bottomrule
  $R^2 = 0.61$
  \end{tabular}
\end{table}

Active time span is measure by the days span from the very initial message to the last message. As members work together longer time, they are more likely to have a better output (marginal significant). We found that the number of bot messages is correlated (but not significantly) to success outcome. As the most used bots across these challenges are the Github bot, it may represent the team is more active in Github related activites that has a better outcome. 

Other features are easy to understand: the result suggests that if there is a single user publishes a lot of the messages in a channel, the more conversation threads created in the channel, the more programming codes, the more emojis, and the more \@here messages are used, the more likely the team has a paper submission. \change{This is the first step towards a promising research future that organizations and managements may leverage only the meta data of group communication practices to predict a team's performance.}




\section{Discussion}
In this section, we describe the the implications of our results. We contribute to updating the understanding of communication styles and in various context-specific  categories.  \change{We trained a machine learning model 
that can automatically categorize Slack group channels in the workplace at reasonably high accuracy. By looking closely at a subset of data with 57 teams that have both Slack communication data and project performance data, we further discuss how this communication style feature metric could be useful for modeling the team performance.}

\subsection{Updating The Understanding of Human-Human Conversation Pattern} 

\change{Through our manual coding, we are able to identify 9 different categories and their distinct communication patterns. This finding updates the existing knowledge of how groups communicate in IM systems. For example, ~\cite{mcdaniel1996identifying} in 1998 reported 156.7 words per thread, whereas in our study we found that number differed in different categories of Slack channels: e.g., \textit{Project} channels have 32.8 words per thread (16.4 words per message * 2.0 messages per thread); but in \textit{IT Support} channels, there have 94.3 words per thread (23.0 words per message * 4.1 messages per thread). This result reflects a dramatic change in people's behaviors after decades of IM technology adoption, when compared to the early adopters in 1998. There are many other aspects of the updated knowledge. Readers can reference to early day's CSCW publications (e.g., ~\cite{nardi2000interaction,mcdaniel1996identifying}), while reading our finds. 

Based on 4,300 data points and the 59 feature metrics we extracted from the meta information, we also built an ML model to automatically classify a group chat channel's category without looking at the actual message content.} We suggest the HCI and NLP researchers who are doing content analysis on group chat message data corpus may first use our model to categorize their group chat channels before building any downstream applications. The current practices of treating all types of group channels equally and using one single setting for all the categories may be misleading \change{(e.g., ~\cite{zhang2018making} investigates how to generate summarizations for group chat messages). }

\subsection{Actionable Insights for NLP Algorithm and AI System Design} 
Also, for NLP researchers who are building deep learning models with synthetic datasets for group chats, they should consider the particular domain and the context that their target audience is in, and refer to our Table~\ref{tab:feat_full} to construct the data corpus that follows a natural distribution. \change{For example, ~\cite{wang2019extracting} builds a deep learning model to automatically extract threads from un-threaded group chat channels with a synthetic Reddit dataset that assumed a thread of conversation should consists of 10 to 100 messages, whereas our results show that on average a thread has 2 to 35 messages. }


\change{One particular active research topic in IM systems today is the AI-powered conversational agent systems~\cite{ashktorab2019resilient,liao2018all}. For example, ~\cite{liao2018all} built and tested an HR bot to support the new employee onboarding process. But these chatbots often rely on dialogue acts or manually crafted dialogue trees, but rarely consider the group conversation context in which the chatbot will be deployed. Using our work's findings, researchers and developers can also consider the contextual norm of human group chat channels. Thus, for different categories of channels, the chatbot can behave differently. For example, when a chatbot is for an Event organization group, it may use much fewer threaded conversations than in an IT Support group. That is how an intelligent conversational system can better fit into the human-human communication group.}

We also envision that the group communication feature metrics could be use to predict team performance in the future. Though not every feature is significant in the regression model, the goodness of the full model ($R^2=0.61$) hints at the promising future of this line of research. If we can build a dashboard or a system that actively track the group conversations in their team channels of a project team, we may be able to have a real time meter for the program team's performance, and an early sign of project failure could be detected. 



\subsection{Limitations}
Our study has a couple limitations. First, the context is within a R\&D department of a multinational IT company, and we use a publication as a proxy for success in this study. It is important to note that success can be defined in other ways in projects (patents, product impact). However, within the context of this time-bounded re-organization even in the R\&D department, it is sufficient to use publications as a proxy for success. Thus, the reader should be warned that some of the results from this study, such as what communication styles lead to higher group performance, may not be generalizable to other contexts.

Secondly, the pre-trained machine learning model for category identification may not generalize well for group chat dataset other than Slack, and there will be a need for retraining of the model on the new dataset to fit for a different feature distribution.

\section{Conclusion}
In this short note, we provide a comprehensive set of three analyses on understanding communication styles in Slack group chat channels in today's workplace settings. We first manually coded 9 different categories of group categories. \change{Then, based on a communication style metric with 21 metadata features, we built a machine learning model that can automatically categorize group chat channels. Finally, we illustrated that these features  could be used to unveil the relation between communication styles and the success of a project team.}

\begin{acks}
We thank all the reviewers for their valuable revision comments. We particularly thank Stacy F Hobson and Talia Gershon for their help and support.
\end{acks}

%
\bibliographystyle{ACM-Reference-Format}
\bibliography{sample-base}

%
\newpage
\appendix
\section{Appendix A: The Feature Metrics For The 9 Categories}

\begin{table}[htp]
\small
  \centering
  \caption{The Feature Metrics For The 9 Categories Identified by Manually Coding 100 Channels Randomly Selected from 4,300 Public Slack Channels in A Company}
  \label{tab:feat_full}
  \begin{tabular}{p{2.5cm} |p{0.9cm}|p{0.9cm}|p{0.9cm}|p{0.9cm}|p{0.9cm}|p{0.9cm}|p{0.9cm}|p{0.9cm}|p{0.9cm}}
    \toprule
    \textbf{Feature} & \textbf{Anno- uncement N=2} & \textbf{Bot N=8} & \textbf{Emplo- yee Support N=3} & \textbf{Event N=9} & \textbf{IT Support N=8} & \textbf{Project N=32} & \textbf{Social Group N=10} & \textbf{Tech Enthusiasts N=10} & \textbf{Test N=18} \\
    \midrule
\#members & 937.0 & 60.0 & 78.3 & 32.8 & 177.0 & 11.1 & 84.0 & 87.6 & 2.7 \\
\#active\_users & 117.0 & 72.5 & 70.3 & 34.1 & 139.9 & 14.4 & 92.7 & 70.8 & 5.4 \\
active\_timespan & 633.5 & 443.8 & 290.7 & 144.1 & 506.5 & 281.2 & 478.6 & 411.9 & 200.7 \\
\#messages & 390.0 & 5908.1 & 267.7 & 54.2 & 4537.4 & 355.9 & 3059.0 & 659.3 & 7.9 \\
word\_per\_message & 20.7 & 18.5 & 20.8 & 9.8 & 23.0 & 16.4 & 12.8 & 17.1 & 5.8 \\
\#messages\_top\_user & 63.5 & 285.9 & 52.0 & 10.4 & 1819.9 & 90.2 & 554.9 & 83.7 & 2.4 \\
\#at\_messages & 209.0 & 305.6 & 108.7 & 39.2 & 953.2 & 79.2 & 337.6 & 217.7 & 6.0 \\
\#channel\_messages & 0.0 & 0.5 & 0.0 & 0.0 & 0.1 & 0.0 & 0.1 & 0.2 & 0.0 \\
\#here\_messages & 0.0 & 47.5 & 0.7 & 0.1 & 1.6 & 0.2 & 8.1 & 0.3 & 0.0 \\
\#thread\_messages & 75.5 & 42.4 & 15.3 & 1.4 & 185.8 & 15.3 & 152.5 & 53.4 & 0.1 \\
max\_turn\_thread & 6.5 & 4.1 & 6.3 & 1.2 & 34.9 & 6.7 & 28.6 & 9.6 & 0.1 \\
avg\_turn\_thread & 1.3 & 0.4 & 2.8 & 0.8 & 4.1 & 2.0 & 2.8 & 2.2 & 0.1 \\
max\_reaction\_count & 15.5 & 4.2 & 9.3 & 2.9 & 9.6 & 2.3 & 13.6 & 9.1 & 0.1 \\
avg\_reaction\_count & 4.5 & 0.5 & 2.2 & 1.5 & 1.5 & 0.9 & 2.2 & 1.5 & 0.1 \\
\#pinned\_messages & 3.0 & 3.2 & 1.0 & 1.0 & 5.5 & 0.4 & 1.9 & 2.3 & 0.0 \\
\#emoji\_messages & 13.5 & 284.8 & 20.0 & 1.8 & 102.4 & 12.0 & 284.7 & 22.9 & 0.3 \\
\#code\_messages & 0.0 & 0.4 & 0.3 & 0.2 & 81.8 & 6.6 & 1.4 & 2.6 & 0.0 \\
\#url\_messages & 14.5 & 559.8 & 57.0 & 6.1 & 329.5 & 34.2 & 188.8 & 80.0 & 0.2 \\
\#git\_messages & 0.5 & 0.5 & 0.0 & 0.1 & 59.6 & 9.0 & 0.7 & 20.6 & 0.0 \\
\#file\_messages & 7.0 & 3345.0 & 17.3 & 2.1 & 58.9 & 60.4 & 180.5 & 25.1 & 0.1 \\
\#bot\_messages & 0.0 & 3477.0 & 2.0 & 0.0 & 1396.5 & 52.2 & 64.6 & 0.3 & 0.1 \\
  \bottomrule
  \end{tabular}
\end{table}

\received{January 2021}
\received[revised]{July 2021}
\received[accepted]{November 2021}

\end{document}